\documentclass[
   aip,
   adv,
   twocolumn,
   reprint,
   superscriptaddress,
   floatfix,
   citeautoscript,
   longbibliography,
]{revtex4-2}

\usepackage[T1]{fontenc}
\usepackage[utf8]{inputenc}

\usepackage{newtxtext}
\usepackage{newtxmath}

\usepackage{chemformula}
\usepackage{siunitx}
\DeclareSIUnit{\ppma}{ppma}

\usepackage{graphicx}
\graphicspath{{./figures}}

\usepackage[
   unicode,
   colorlinks = true,
   allcolors = blue,
]{hyperref}

\usepackage{cleveref}
\usepackage{microtype}

\usepackage{glossaries}
\glsdisablehyper

\newacronym{le-musr}{LE-\ensuremath{\mu}SR}{low-energy muon spin rotation}
\newacronym{ltb}{LTB}{low-temperature baking}
\newacronym{srf}{SRF}{superconducting radio frequency}
\newacronym{psi}{PSI}{Paul Scherrer Institute}
\newacronym{sms}{S\ensuremath{\mu}S}{Swiss Muon Source}
\newacronym{sims}{SIMS}{secondary ion mass spectrometry}
\newacronym{bcp}{BCP}{buffered chemical polishing}
\newacronym{ep}{EP}{electro-polishing}
\newacronym{hfqs}{HFQS}{high-field \ensuremath{Q} slope}
\newacronym{ss}{SS}{superconductor-superconductor}
\newacronym{nserc}{NSERC}{Natural Sciences and Engineering Research Council of Canada}

\DeclareMathOperator{\erfc}{erfc}

\makeatletter
\def\@email#1#2#3{%
 \endgroup
 \patchcmd{\titleblock@produce}
  {\frontmatter@RRAPformat}
  {\frontmatter@RRAPformat{\produce@RRAP{*#1\href{mailto:#2}{#2}; and \href{mailto:#3}{#3}}}\frontmatter@RRAPformat}
  {}{}
}%
\makeatother

\begin{document}

\title{
	Search for inhomogeneous Meissner screening in \ch{Nb} induced by
	low-temperature surface treatments
}

% coauthors
\author{Ryan~M.~L.~McFadden}
% \email[E-mail: ]{rmlm@triumf.ca}
\email[Authors to whom correspondence should be addressed: ]{rmlm@triumf.ca}{junginger@uvic.ca}
\affiliation{TRIUMF, 4004 Wesbrook Mall, Vancouver, BC V6T~2A3, Canada}
\affiliation{Department of Physics and Astronomy, University of Victoria, 3800 Finnerty Road, Victoria, BC V8P~5C2, Canada}

\author{Tobias~Junginger}
% \email[E-mail: ]{junginger@uvic.ca}
\affiliation{TRIUMF, 4004 Wesbrook Mall, Vancouver, BC V6T~2A3, Canada}
\affiliation{Department of Physics and Astronomy, University of Victoria, 3800 Finnerty Road, Victoria, BC V8P~5C2, Canada}

% today's date
\date{\today}

\begin{abstract}
Empirical surface treatments,
such as \gls{ltb} in a gaseous atmosphere or in vacuum,
are important for the surface preparation of \ch{Nb} \gls{srf} cavities.
These treatments inhomogeneously dope the first \qty{\sim 50}{\nano\meter}
of \ch{Nb}'s subsurface
and are expected to impart depth-dependent characteristics to
its Meissner response;
however, direct evidence supporting this remains elusive,
suggesting the effect is subtle.
In this work,
we revisit the Meissner profile data for several \gls{ltb} treatments
obtained from \gls{le-musr} experiments
[A.~Romanenko \emph{et al}., \href{https://doi.org/10.1063/1.4866013}{Appl.\ Phys.\ Lett.\ \textbf{104}, 072601 (2014)}
and
R.~M.~L.~McFadden \emph{et al}., \href{https://doi.org/10.1103/PhysRevApplied.19.044018}{Phys.\ Rev.\ Appl.\ \textbf{19}, 044018 (2023)}],
and search for signatures of inhomogeneous field screening.
Using a generalized London expression with a recently proposed empirical model
for a depth-dependent magnetic penetration depth $\lambda(z)$,
we obtain improved fits to the Meissner data,
revealing that the presence of a non-superconducting surface ``dead layer''
$d \gtrsim \qty{25}{\nano\meter}$ is a strong indicator of a reduced supercurrent
density at shallow subsurface depths.
Our analysis supports the notion that vacuum annealing at \qty{120}{\celsius}
for \qty{48}{\hour} induces a depth-dependent Meissner response,
which has consequences for \ch{Nb}'s ability to maintain a magnetic-flux-free state.
Evidence of similar behavior from a ``nitrogen infusion'' treatment is less compelling.
Suggestions for further investigation into the matter are provided.
\end{abstract}

\maketitle
\glsresetall

\section{Introduction}

\Gls{ltb} treatments are important for the surface preparation
of \ch{Nb} \gls{srf} cavities~\cite{2008-Padamsee-RFSA-2,2009-Padamsee-RFSSTA,2023-Padamsee-SRTA}
---
a class of superconducting resonators~\cite{2023-Gurevich-SST-36-063002} used in particle accelerators.
These procedures,
performed either in vacuum~\cite{2004-Ciovati-JAP-96-1591,arXiv:1806.09824}
or
in a low-pressure gaseous environment~\cite{2013-Grassellino-SST-26-102001,2017-Grassellino-SST-30-094004},
induce changes to \ch{Nb}'s supercurrent density in its near-surface region
through the introduction of intrinsic
(e.g., dissolved oxygen from its surface oxide layer~\cite{2004-Ciovati-JAP-96-1591})
or
extrinsic
(e.g., infused nitrogen from the treatment atmosphere~\cite{2017-Grassellino-SST-30-094004})
shallow impurity profiles.
Chiefly,
these treatments are known to mitigate the \gls{hfqs} in \gls{srf} cavities~\cite{1999-Lilje-WoRFS-9-74,2013-Romanenko-PRSTAB-16-012001,2018-Dhakal-PRAB-21-032001,2020-Checchin-APL-117-032601},
wherein a sudden drop in a cavity's quality factor $Q$ is coincident with
a sharp increase in its surface resistance~\cite{2023-Padamsee-SRTA}.
Their mechanism of action is thought to be connected to the nanoscale
alternations taking place in \ch{Nb}'s near-surface region
(i.e., the topmost \qty{\sim 50}{\nano\meter}),
where
empirical observations~\cite{2013-Romanenko-PRSTAB-16-012001,2017-Grassellino-SST-30-094004,2020-Checchin-APL-117-032601}
and
model predictions~\cite{2006-Ciovati-APL-89-022507,2019-Kubo-PRB-100-064522,2024-Lechner-JAP-135-133902}
suggest that the treatment-induced defects are \emph{inhomogeneous} with depth.
Of particular interest is how this impacts \ch{Nb}'s ability
to remain in a flux-free Meissner state~\cite{2017-Junginger-SST-30-125012,2018-Junginger-PRAB-21-032002,2021-Kubo-SST-34-045006,2022-Turner-SR-12-5522,2024-Asaduzzaman-SST-37-085006},
which is essential for \gls{srf} cavity operation~\cite{2023-Padamsee-SRTA}.
As several material properties are proportional to impurity concentration
(e.g., carrier mean-free-path $\ell$~\cite{1968-Goodman-JPF-29-240,1976-Schulze-ZM-67-737}),
they too are expected to vary non-uniformly below the surface,
which should influence the element's
Meissner response~\cite{2019-Ngampruetikorn-PRR-1-012015,2020-Checchin-APL-117-032601,2024-Lechner-JAP-135-133902};
however,
directly probing their influence
remains challenging.

Early measurements on a
\ch{Nb}
\gls{srf} cavity cutout baked at \qty{120}{\celsius}
for \qty{48}{\hour}
(which we refer to henceforth as ``\qty{120}{\celsius} bake'') 
showed an abrupt
discontinuity in its Meissner profile~\cite{2014-Romanenko-APL-104-072601},
suggesting the presence of a spatially inhomogeneous near-surface supercurrent density.
Among the initial efforts to model this behavior
(see, e.g., Refs.~\citenum{2016-Checchin-PhD,2016-Checchin-IPAC-7-2254}),
one approach~\cite{2017-Kubo-SST-30-023001},
in analogy with superconducting multilayers~\cite{2006-Gurevich-APL-88-012511,2014-Kubo-APL-104-032603,2015-Gurevich-AIPA-5-017112,2015-Posen-PRA-4-044019},
approximated the inhomogeneous surface region as a \gls{ss} bilayer
(i.e., a thin ``dirty'' layer atop a ``clean'' substrate),
correctly reproducing the treatment's
abated near-surface supercurrent density~\cite{2017-Kubo-SST-30-023001}.
Despite this breakthrough,\footnote{This finding in Ref.~\citenum{2017-Kubo-SST-30-023001} is noteworthy, as it demonstrates how inhomogeneous screening properties lead to the suppression of surface supercurrents, which in turn increase \ch{Nb}'s superheating field $B_{\mathrm{sh}}$ (i.e., the maximum magnetic field to which the metal can remain in a (metastable) flux-free state) --- an operational limit realized by state-of-the-art \gls{srf} cavities (see, e.g., Refs.~\citenum{2023-Padamsee-SRTA,2023-Gurevich-SST-36-063002}).}
subsequent
measurements on a similarly treated
\ch{Nb} sample showed no such feature~\cite{2023-McFadden-PRA-19-044018},
with
true
bilayer samples also producing distinct screening properties~\cite{2024-Asaduzzaman-SST-37-025002}.
These differences were explained in a recent commentary~\cite{2024-McFadden-APL-124-086101},
with the discontinuity attributed to an analysis artifact.
Interestingly,
this work also uncovered a large non-superconducting surface
``dead layer'' $d \approx \qty{35}{\nano\meter}$~\cite{2024-McFadden-APL-124-086101}.
While $d \gtrsim \qty{5}{\nano\meter}$ for \ch{Nb} is
common~\cite{2005-Suter-PRB-72-024506,2014-Romanenko-APL-104-072601,2017-Junginger-SST-30-125013,2023-McFadden-PRA-19-044018,2023-McFadden-JAP-134-163902},
likely due to the roughness~\cite{2014-Lindstrom-JEM-85-149,2016-Lindstrom-JSNM-29-1499}
introduced by its surface oxidation~\cite{1987-Halbritter-APA-43-1},
such a large value is suggestive of a surface-localized region
with a diminished screening capacity.
This finding is consistent with the shallow impurity profiles found
in closely related \gls{ltb} treatments~\cite{2019-Romanenko-SRF-19-866,2021-Lechner-APL-119-082601,2024-Lechner-JAP-135-133902},
which are predicted to ``distort'' the Meissner profile close to the surface
(see, e.g., Ref.~\citenum{2024-Lechner-JAP-135-133902}).
The effect
is,
however,
likely
both gradual and subtle~\cite{2024-Lechner-JAP-135-133902,2024-McFadden-APL-124-086101},
requiring scrupulous inspection.

\begin{figure}
	\centering
	\includegraphics[width=1.0\columnwidth]{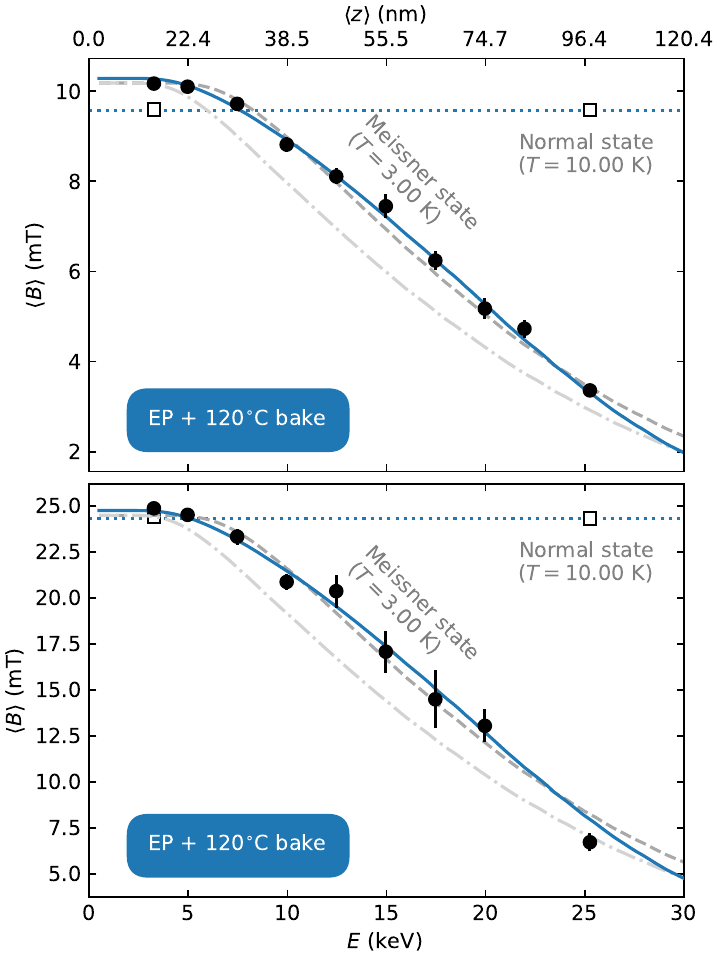}
	\caption{
		\label{fig:mcfadden-2024}
		Meissner screening data
		(at two different applied fields)
		in a \ch{Nb} \gls{srf} cavity cutout ``cold spot''
		with an ``\gls{ep}~\cite{2011-Ciovati-JAE-41-721} + \qty{120}{\celsius} bake~\cite{2004-Ciovati-JAP-96-1591}'' surface treatment,
		obtained from \gls{le-musr} experiments originally reported in
		Ref.~\citenum{2014-Romanenko-APL-104-072601}
		(and recently re-analyzed in
		Ref.~\citenum{2024-McFadden-APL-124-086101}).
		Here,
		the mean magnetic field $\langle B \rangle$ is plotted against
		the muon $\mu^{+}$ implantation energy $E$,
		with its corresponding mean stopping depth $\langle z \rangle$
		indicated on a secondary axis.
		The solid and dotted colored lines denote simultaneous fits of the
		normal and Meissner state data to
		\Cref{eq:average-field,eq:london-inhomogeneous,eq:lambda-inhomogeneous}.
		For comparison,
		the dashed dark gray line shows the form of $\langle B \rangle (E)$
		assuming a single (i.e., depth-independent) magnetic penetration depth
		$\lambda$
		[\Cref{eq:average-field,eq:london,eq:effective-field}].
		The dash-dot light gray line denotes this average screening behavior,
		but with its non-superconducting ``dead layer'' $d$ adjusted to match the fit
		using the generalized London equation,
		highlighting the differences between the two models.
	}
\end{figure}

\begin{figure}
	\centering
	\includegraphics[width=1.0\columnwidth]{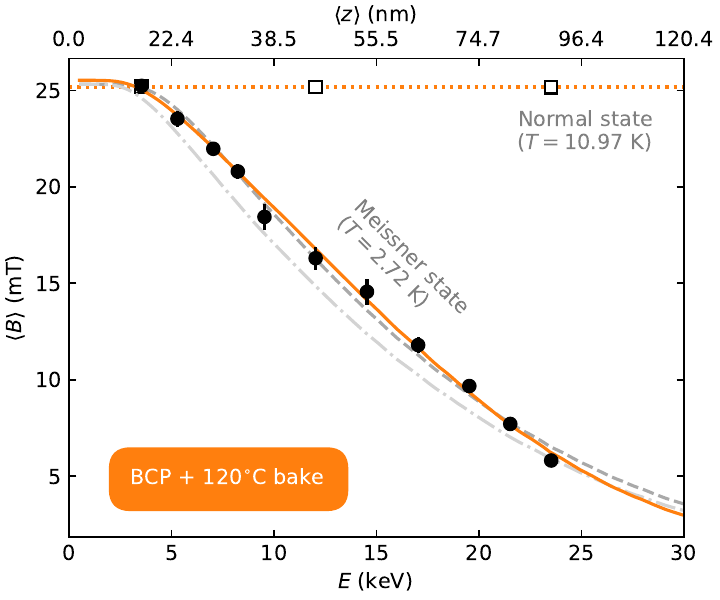}
	\caption{
		\label{fig:mcfadden-2023-ltb}
		Meissner screening data in an \gls{srf}-grade \ch{Nb} plate with
		a ``\gls{bcp}~\cite{2011-Ciovati-JAE-41-721} + \qty{120}{\celsius} bake~\cite{2004-Ciovati-JAP-96-1591}'' surface treatment,
		obtained from \gls{le-musr} experiments originally reported in
		Ref.~\citenum{2023-McFadden-PRA-19-044018}.
		Here,
		the mean magnetic field $\langle B \rangle$ is plotted against
		the muon $\mu^{+}$ implantation energy $E$,
		with its corresponding mean stopping depth $\langle z \rangle$
		indicated on a secondary axis.
		The solid and dotted colored lines denote simultaneous fits of the
		normal and Meissner state data to
		\Cref{eq:average-field,eq:london-inhomogeneous,eq:lambda-inhomogeneous}.
		For comparison,
		the dashed dark gray line shows the form of $\langle B \rangle (E)$
		assuming a single (i.e., depth-independent) magnetic penetration depth
		$\lambda$
		[\Cref{eq:average-field,eq:london,eq:effective-field}].
		The dash-dot light gray line denotes this average screening behavior,
		but with its non-superconducting ``dead layer'' $d$ adjusted to match the fit
		using the generalized London equation,
		highlighting the differences between the two models.
	}
\end{figure}

\begin{figure}
	\centering
	\includegraphics[width=1.0\columnwidth]{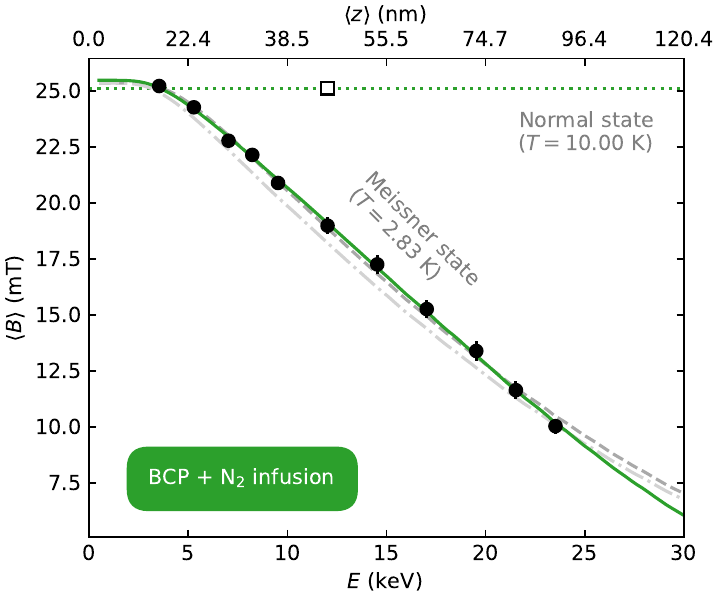}
	\caption{
		\label{fig:mcfadden-2023-n2i}
		Meissner screening data in an \gls{srf}-grade \ch{Nb} plate with
		a ``\gls{bcp}~\cite{2011-Ciovati-JAE-41-721} +
		\ch{N2} infusion''~\cite{2017-Grassellino-SST-30-094004} surface treatment,
		obtained from \gls{le-musr} experiments originally reported in
		Ref.~\citenum{2023-McFadden-PRA-19-044018}.
		Here,
		the mean magnetic field $\langle B \rangle$ is plotted against
		the muon $\mu^{+}$ implantation energy $E$,
		with its corresponding mean stopping depth $\langle z \rangle$
		indicated on a secondary axis.
		The solid and dotted colored lines denote simultaneous fits of the
		normal and Meissner state data to
		\Cref{eq:average-field,eq:london-inhomogeneous,eq:lambda-inhomogeneous}.
		For comparison,
		the dashed dark gray line shows the form of $\langle B \rangle (E)$
		assuming a single (i.e., depth-independent) magnetic penetration depth
		$\lambda$
		[\Cref{eq:average-field,eq:london,eq:effective-field}].
		The dash-dot light gray line denotes this average screening behavior,
		but with its non-superconducting ``dead layer'' $d$ adjusted to match the fit
		using the generalized London equation,
		highlighting the differences between the two models.
	}
\end{figure}

In this work,
we revisit measurements of \ch{Nb}'s Meissner screening profile
obtained by \gls{le-musr}~\cite{2004-Bakule-CP-45-203,2004-Morenzoni-JPCM-16-S4583}
---
a sensitive, non-destructive technique for interrogating subsurface magnetic
fields with nanometer depth resolution
---
for several \gls{ltb} treatments
and
search for evidence of spatially inhomogeneous screening.
Specifically,
we examine results obtained for:
a \ch{Nb} \gls{srf} cavity ``cold spot'' cutout whose surface underwent
\gls{ep}~\cite{2011-Ciovati-JAE-41-721}
+
\qty{120}{\celsius} baking~\cite{2004-Ciovati-JAP-96-1591}
(see Refs.~\citenum{2014-Romanenko-APL-104-072601,2024-McFadden-APL-124-086101}
and \Cref{fig:mcfadden-2024});
an \gls{srf}-grade \ch{Nb} plate with a surface prepared by
\gls{bcp}~\cite{2011-Ciovati-JAE-41-721} +
\qty{120}{\celsius} baking~\cite{2004-Ciovati-JAP-96-1591}
(see Ref.~\citenum{2023-McFadden-PRA-19-044018}
and \Cref{fig:mcfadden-2023-ltb,fig:mcfadden-2023-all});
and
another \gls{srf} \ch{Nb} plate with a surface prepared by
\gls{bcp}~\cite{2011-Ciovati-JAE-41-721}
+
``\ch{N2} infusion''~\cite{2017-Grassellino-SST-30-094004}
(see Ref.~\citenum{2023-McFadden-PRA-19-044018}
and \Cref{fig:mcfadden-2023-n2i,fig:mcfadden-2023-all}).
By comparing the form of the each sample's screening profile against a recently
proposed phenomenological model~\cite{2020-Checchin-APL-117-032601},
we search for signatures of inhomogeneous screening.

\section{Experimental Background \& Theory}

In \gls{le-musr}~\cite{2004-Bakule-CP-45-203,2004-Morenzoni-JPCM-16-S4583},
positive muons $\mu^{+}$
(spin $S = 1/2$;
gyromagnetic ratio $\gamma / (2 \pi) = \qty{135.539}{\mega\hertz\per\tesla}$;
mean lifetime $\tau = \qty{2.197}{\micro\second}$)
are implanted in a host material at an energy $E$ between
\qtyrange{0.5}{30}{\kilo\electronvolt},
where they probe subsurface electromagnetic fields at depths up to
\qty{\sim 200}{\nano\meter}~\cite{2002-Morenzoni-NIMB-192-245}.
The technique is highly sensitive to spatial gradients in magnetic field,
making it well-suited for the study of Meissner profiles~\cite{2004-Morenzoni-JPCM-16-S4583}.
In the measurement,
one observes the $\mu^{+}$ spin-precession signal
(detected via its radioactive decay products),
which provides a measure of the field distribution $p(B)$
sampled by the $\mu^{+}$ stopping profile $\rho(z, E)$~\footnote{For typical $\mu^{+}$ stopping profiles in \ch{Nb} at $E \leq \qty{30}{\kilo\electronvolt}$, see Fig.~1 in Ref.~\citenum{2023-McFadden-PRA-19-044018}.},
where $z$ denotes the depth below the surface.
Specific experimental details can be found in Refs.~\citenum{2014-Romanenko-APL-104-072601,2023-McFadden-PRA-19-044018,2024-Asaduzzaman-SST-37-025002},
with broader technique overviews provided elsewhere~\cite{2021-Blundell-MSI,2022-Hillier-NRMP-2-4,2024-Amato-IMSS}.

In the measurements,
it is useful to identify the average field 
$\langle B \rangle \equiv \int_{-\infty}^{\infty} p(B) B \, \mathrm{d}B$
as a function of $E$,
the latter being proportional to the mean $\mu^{+}$ stopping depth $\langle z \rangle \equiv \int_{0}^{\infty} z \rho(z, E) \, \mathrm{d} z$.
$\langle B \rangle$
is related to the \emph{true} screening profile $B(z)$ by:
\begin{equation}
	\label{eq:average-field}
	\langle B \rangle (E) = \int_{0}^{\infty} B(z) \rho(z, E) \, \mathrm{d}z ,
\end{equation}
where $\rho(z,E)$ acts as the kernel for the integral
transform~\footnote{Note that $\rho(z,E)$ is not directly determined by the measurement; however, it can be accurately quantified using Monte Carlo simulations for $\mu^{+}$ implantation~\cite{2002-Morenzoni-NIMB-192-245,2023-McFadden-PRA-19-044018}. While the simulation results are generally obtained at discrete $E$, they can be made continuous using empirical ``interpolation'' schemes~\cite{2023-McFadden-PRA-19-044018,2024-Asaduzzaman-SST-37-025002}, facilitating the evaluation of \Cref{eq:average-field}.}.
In previous work~\cite{2014-Romanenko-APL-104-072601,2023-McFadden-PRA-19-044018,2024-McFadden-APL-124-086101},
$B(z)$ was found to be consistent with a London model~\cite{1935-London-PRSLA-149-71},
following:
\begin{equation}
	\label{eq:london}
	B(z) = \tilde{B}_{0} \times \begin{cases}
		1, & z < d , \\
		\displaystyle \exp \left \{ -\frac{ ( z - d ) }{ \lambda } \right \} , & z \geq d , 
	\end{cases}
\end{equation} 
where
$\lambda$ is the
(spatially homogeneous)
magnetic penetration depth,
$d$ is ``dead layer'' thickness,
and
$\tilde{B}_{0}$ is the (effective) applied magnetic field:
\begin{equation}
	\label{eq:effective-field}
	\tilde{B}_{0} = B_{\mathrm{applied}} \times \begin{cases}
		1, & T \geq T_{c} , \\
		\left ( 1 - \tilde{N} \right )^{-1} , & T \ll T_{c} , \\
	\end{cases}
\end{equation}
where $B_{\mathrm{applied}}$ is the applied magnetic field,
$T_{c} \approx \qty{9.25}{\kelvin}$ is \ch{Nb}'s critical temperature~\cite{2022-Turner-SR-12-5522},
and
$\tilde{N}$ is the sample's (effective) demagnetization factor
(i.e., averaged over the $\mu^{+}$ beam profile).
While this gave a good description of the measurements,
the single depth-independent $\lambda$ in \Cref{eq:london} is inconsistent with
predictions by
others~\cite{2020-Checchin-APL-117-032601,2024-Lechner-JAP-135-133902}.

To account for spatial inhomogeneities in the Meissner profile,
we consider the following generalization of the one-dimension London
equation~\cite{1935-London-PRSLA-149-71,1981-Simon-PRB-23-4463,1986-Cave-JLTP-63-35,1994-Pambianchi-PRB-50-13659}:
\begin{equation}
	\label{eq:london-inhomogeneous}
	\lambda^{2}(z) \left [ \frac{ \mathrm{d}^{2} B(z) }{ \mathrm{d}z^{2}} \right ] + 2 \lambda(z) \left [ \frac{ \mathrm{d} \lambda(z) }{ \mathrm{d}z } \right ] \left [ \frac{ \mathrm{d} B(z) }{ \mathrm{d}z } \right ] = B(z) ,
\end{equation}
where $B(z)$ is the magnetic field at depth $z$,
and
$\lambda(z)$ is the depth-dependent magnetic penetration
depth~\footnote{Note that when $\mathrm{d}\lambda(z)/\mathrm{d}z = 0$ (i.e., when $\lambda(z)$ is depth-independent), \Cref{eq:london-inhomogeneous} reduces to the familiar London expression~\cite{1935-London-PRSLA-149-71}, whose solution $B(z) \propto \exp(-z / \lambda)$ is akin to \Cref{eq:london}.}.
This expression is often encountered in situations where a spatial dependence to
$\lambda$ is anticipated,
such as in proximity-coupled superconductor/normal-metal interfaces
(see, e.g., Refs.~\citenum{1981-Simon-PRB-23-4463,1994-Pambianchi-PRB-50-13659}).
Note that analytic solutions to \Cref{eq:london-inhomogeneous}
can only be obtained for select models of $\lambda(z)$~\cite{1994-Pambianchi-PRB-50-13659},
with the general case requiring numerical methods~\cite{1986-Cave-JLTP-63-35}.
In the context of \gls{srf} cavities,
it has been proposed that $\lambda(z)$ has the form~\cite{2020-Checchin-APL-117-032601}:
\begin{equation}
	\label{eq:lambda-inhomogeneous}
	\lambda(z) = \left ( \lambda_{\mathrm{surface}} - \lambda_{\mathrm{bulk}} \right ) \erfc \left ( \frac{z}{L_{D}} \right )  + \lambda_{\mathrm{bulk}} ,
\end{equation}
where the $\lambda_{i}$s denote the penetration depths at \ch{Nb}'s surface and bulk, respectively,
with $\lambda(z)$ interpolating the two values over a length scale $L_{D}$
(e.g., the diffusion length of impurity atoms~\cite{2020-Checchin-APL-117-032601,2021-Lechner-APL-119-082601,2024-Lechner-JAP-135-133902})
according to the complimentary error function
$\erfc (x) \equiv 1 - (2 / \pi ) \int_{0}^{x} \exp ( -y^{2} ) \, \mathrm{d}y $.
Alternative approaches for characterizing $\lambda(z)$ have also been proposed,
producing qualitatively similar $B(z)$s~\cite{2024-Lechner-JAP-135-133902}.

\section{Results \& Discussion}

To quantify the screening behavior explicitly,
we fit the \gls{le-musr} data in
\Cref{fig:mcfadden-2024,fig:mcfadden-2023-ltb,fig:mcfadden-2023-n2i},
using \Cref{eq:average-field}
and numeric solutions to \Cref{eq:london-inhomogeneous},
constraining $\lambda(z)$ to follow \Cref{eq:lambda-inhomogeneous}
with $\lambda_{\mathrm{bulk}} = \qty{29}{\nano\meter}$
(in accord with ``clean'' \ch{Nb}'s literature
average~\cite{2023-McFadden-PRA-19-044018,2023-McFadden-JAP-134-163902}),
and
parameterizing $B(z)$ to contain both a finite $d$ and
a geometrically enhanced field $B_{\mathrm{applied}}/(1 - \tilde{N})$,
akin to \Cref{eq:london,eq:effective-field}.
Fit results are shown in
\Cref{fig:mcfadden-2024,fig:mcfadden-2023-ltb,fig:mcfadden-2023-n2i},
in good agreement with the measurements.
For comparison,
screening profiles assuming a spatially homogenous $\lambda$
[\Cref{eq:average-field,eq:london,eq:effective-field}]
are also shown.
A summary of these results is given in \Cref{tab:results}
and
we consider the details below.

\begin{table*}
	\centering
	\caption{
		\label{tab:results}
		Summary of fit results for the \gls{le-musr} measurements of the
		Meissner profiles in \gls{srf} \ch{Nb} samples with \gls{ltb}
		surface-treatments.
		For each sample,
		its surface-treatment and Meissner profile measurement temperature
		$T$ is indicated,
		along values for the fit parameters obtained using
		\Cref{eq:average-field,eq:london-inhomogeneous,eq:lambda-inhomogeneous}
		(top three rows)
		and
		\Cref{eq:average-field,eq:london,eq:effective-field}
		(bottom three rows).
		Here,
		$B_{\mathrm{applied}}$ is the applied magnetic field,
		$\tilde{N}$ is the (effective) demagnetization factor,
		$d$ is the non-superconducting ``dead layer'' thickness,
		$\lambda$ is the depth-independent magnetic penetration depth,
		$\lambda_{\mathrm{surface}}$ is the magnetic penetration depth at the surface,
		$\lambda_{\mathrm{bulk}}$ is the magnetic penetration depth in the bulk,
		and
		$L_{D}$ is the length scale in \Cref{eq:lambda-inhomogeneous}
		over which $\lambda(z)$ changes from
		$\lambda_{\mathrm{surface}}$ to $\lambda_{\mathrm{bulk}}$.
		In each case,
		the goodness-of-fit metric $\chi_{\mathrm{reduced}}^{2}$
		is also provided.
	}
\begin{tabular*}{\textwidth}{l @{\extracolsep{\fill}} S S S S S S S S S}
\botrule
{Sample} & {$T$ (\si{\kelvin})} & {$B_{\mathrm{applied}}$ (\unit{\milli\tesla})} & {$\tilde{N}$} & {$d$ (\si{\nano\meter})} & {$\lambda$ (\unit{\nano\meter})} & {$\lambda_{\mathrm{surface}}$ (\unit{\nano\meter})} & {$\lambda_{\mathrm{bulk}}$ (\unit{\nano\meter})} & {$L_{D}$ (\unit{\nano\meter})} & {$\chi_{\mathrm{reduced}}^{2}$} \\
\hline
\acrshort{ep} + \qty{120}{\celsius} bake & 3.00 & {\num{9.58 \pm 0.04}\, / \,\num{24.33 \pm 0.06}} & {\num{0.069 \pm 0.011}\, / \,\num{0.017 \pm 0.011}} & 26 \pm 4 &  & 78 \pm 12 & 29 & 63 \pm 9 & 0.91 \\
\acrshort{bcp} + \qty{120}{\celsius} bake & 2.72 & 25.179 \pm 0.034 & 0.014 \pm 0.010 & 20.9 \pm 2.2 &  & 54 \pm 6 & 29 & 57 \pm 12 & 1.75 \\
BCP + \ch{N2} infusion & 2.83 & 25.11 \pm 0.06 & 0.014 \pm 0.010 & 21.3 \pm 2.3 &  & 75 \pm 5 & 29 & 140 \pm 60 & 0.44 \\
\hline
\acrshort{ep} + \qty{120}{\celsius} bake & 3.00 & {\num{9.58 \pm 0.04}\, / \,\num{24.33 \pm 0.06}} & {\num{0.059 \pm 0.009}\, / \,\num{0.007 \pm 0.009}} & 34.9 \pm 1.4 & 52.5 \pm 1.8 &  &  &  & 1.73 \\
\acrshort{bcp} + \qty{120}{\celsius} bake & 2.72 & 25.179 \pm 0.034 & 0.006 \pm 0.011 & 25.4 \pm 1.3 & 42.6 \pm 1.3 &  &  &  & 2.09 \\
BCP + \ch{N2} infusion & 2.83 & 25.11 \pm 0.06 & 0.009 \pm 0.011 & 24.1 \pm 1.7 & 70.2 \pm 2.6 &  &  &  & 0.62 \\

\botrule
\end{tabular*}
\end{table*}

\begin{figure}
	\centering
	\includegraphics[width=1.0\columnwidth]{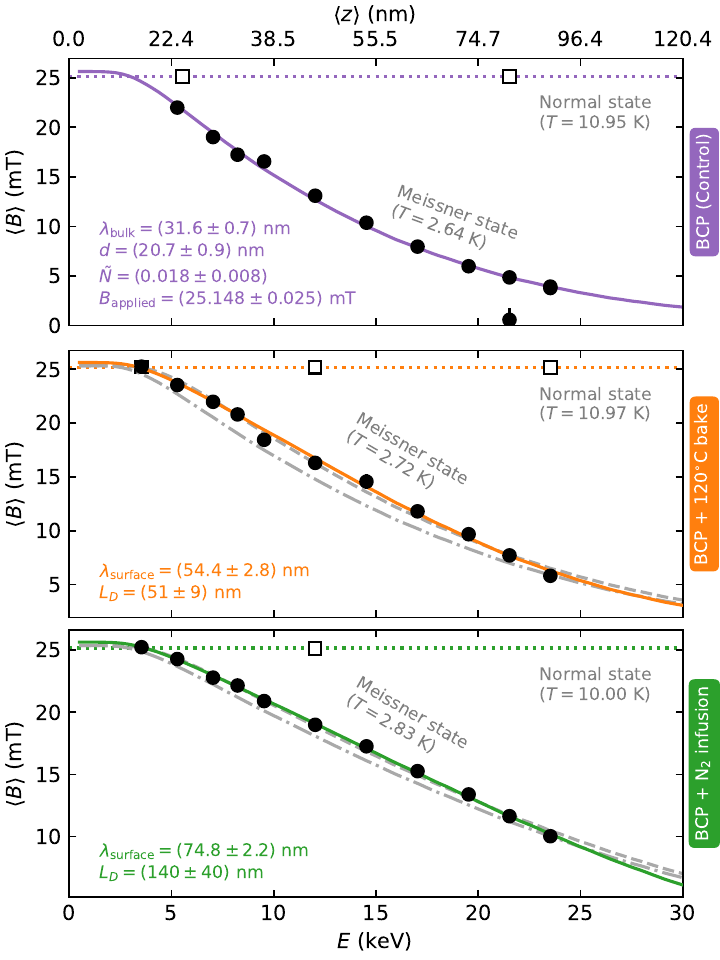}
	\caption{
		\label{fig:mcfadden-2023-all}
		Meissner screening data in a \gls{srf}-grade \ch{Nb} plates
		with ``\gls{bcp}~\cite{2011-Ciovati-JAE-41-721}'' (top panel),
		``\gls{bcp}~\cite{2011-Ciovati-JAE-41-721} + \qty{120}{\celsius} bake~\cite{2004-Ciovati-JAP-96-1591}'' (middle panel),
		and
		``\gls{bcp}~\cite{2011-Ciovati-JAE-41-721} + \ch{N2} infusion''~\cite{2017-Grassellino-SST-30-094004} (bottom panel) surface treatments,
		obtained from \gls{le-musr} experiments originally reported in
		Ref.~\citenum{2023-McFadden-PRA-19-044018}.
		Here,
		the mean magnetic field $\langle B \rangle$ is plotted against
		the muon $\mu^{+}$ implantation energy $E$,
		with its corresponding mean stopping depth $\langle z \rangle$
		indicated on a secondary axis.
		The solid and dotted colored lines denote simultaneous fits of the
		normal and Meissner state data in \emph{all} samples,
		using \Cref{eq:average-field,eq:london,eq:effective-field} for the
		``\gls{bcp}'' treatment (i.e., as a control)
		and
		\Cref{eq:average-field,eq:london-inhomogeneous,eq:lambda-inhomogeneous}
		for the \gls{ltb} samples,
		with the optimum parameters listed in the panel insets.
		For comparison,
		in the middle and bottom panels 
		the dashed dark gray line shows the form of $\langle B \rangle (E)$
		assuming a single (i.e., depth-independent) magnetic penetration depth
		$\lambda$
		[\Cref{eq:average-field,eq:london,eq:effective-field}]
		in the \gls{ltb} sample.
		Similarly,
		the dash-dot light gray line denotes this average screening behavior,
		but with its non-superconducting ``dead layer'' $d$ adjusted to match the fit
		using the generalized London equation,
		highlighting the differences between the two models.
	}
\end{figure}

First,
we remark that the results obtained using the two screening models are quite similar;
both provide the correct qualitative form of $\langle B \rangle (E)$,
yielding virtually identical values for the ``experimental'' parameters
$B_{\mathrm{applied}}$ and $\tilde{N}$
(see \Cref{tab:results}).
Though some quantitative differences are apparent
(e.g., the $E$ or $\langle z \rangle$ at which $\langle B \rangle$ beings to decay),
both models provide a reasonable description of the Meissner profiles,
as evidenced by their goodness-of-fit metric $\chi_{\mathrm{reduced}}^{2}$.
This is expected,
given the applicability of the London model in previous
work~\cite{2014-Romanenko-APL-104-072601,2023-McFadden-PRA-19-044018,2024-McFadden-APL-124-086101};
however,
we find that the inhomogeneous screening model does a better job of capturing
the features of each dataset
(see \Cref{tab:results}).
While this could simply be a result of the fit function's ``extra''
degrees-of-freedom
(i.e., six vs.\ four parameters),
a judicious inspection of the $d$ values suggest otherwise.

The $d$s identified by the inhomogeneous screening model are all systematically
lower than those extracted assuming a homogeneous Meissner response
(see \Cref{tab:results}).
This is most pronounced for the ``\gls{ep} + \qty{120}{\celsius} bake'' treatment,
with the differences visible by eye in \Cref{fig:mcfadden-2024}.
Specifically,
the ``break in'' of $\langle B \rangle$ from the (homogeneous)
London model is much more abrupt than the data,
in contrast to the smoother curvature of the generalized London equation.
In essence,
the London model overcompensates for this discrepancy by lengthening $d$,
which we take as evidence for near-surface inhomogeneous screening.
Similar behavior is also observed for the ``\gls{bcp} + \qty{120}{\celsius} bake''
sample (see \Cref{fig:mcfadden-2023-ltb}),
though the effect is less pronounced.
Surprisingly,
this ``feature'' is virtually absent from the ``\gls{bcp} + \ch{N2} infusion'' sample,
at odds with the results in Ref.~\citenum{2020-Checchin-APL-117-032601}.
It is important to stress that neither treatment~\cite{2004-Ciovati-JAP-96-1591,2017-Grassellino-SST-30-094004}
meaningfully increases the surface oxide thickness
(see, e.g., Ref.~\citenum{2022-Prudnikava-SST-35-065019}).
At this juncture,
we note that the above analysis has only considered the case of local electrodynamics
governing \ch{Nb}'s Meissner response;
however,
\emph{nonlocal} effects~\cite{1953-Pippard-PRSLA-216-547,1957-Bardeen-PR-108-1175}
are also known to produce diminished screening
close to the surface
(see, e.g., Ref.~\citenum{2005-Suter-PRB-72-024506}).
While nonlocal effects are most pronounced in the ``clean'' limit,
they are known to be weak for \ch{Nb}~\cite{2005-Suter-PRB-72-024506},
and
on the basis of the impure nature of the sample surfaces
(i.e., from their \gls{ltb} treatments)~\cite{2006-Ciovati-APL-89-022507,2017-Grassellino-SST-30-094004,2024-Lechner-JAP-135-133902},
we rule out their importance.
Thus,
we take the differences in the $d$s identified by each model,
in conjunction with the overall fit quality $\chi_{\mathrm{reduced}}^{2}$,
as evidence supporting inhomogeneous near-surface field
screening~\footnote{As a check of this conclusion, we also compared fits of the homogeneous and inhomogeneous screening models to ``clean'' \ch{Nb} with a \gls{bcp} surface (i.e., the ``baseline'' sample in Ref.~\citenum{2023-McFadden-PRA-19-044018}), finding no meaningful differences.}.

Further evidence supporting this notion
can be gleaned from the remaining fit parameters.
For the two ``\qty{120}{\celsius} bake''~\cite{2004-Ciovati-JAP-96-1591} samples,
despite their different surface polishings
(i.e., \gls{ep} vs. \gls{bcp})~\cite{2011-Ciovati-JAE-41-721},
we find that their $\lambda_{\mathrm{surface}}$ and $L_{D}$ values are similar,
suggesting the observed screening behavior is intrinsic for the treatment.
Taking their weighted average,
we find
$L_{D} = \qty{61 \pm 7}{\nano\meter}$,
which is comparable to the spatial extent of the oxygen impurity profile
induced by baking~\cite{2006-Ciovati-APL-89-022507,2013-Romanenko-PRSTAB-16-012001,2024-Lechner-JAP-135-133902},
and
$\lambda_{\mathrm{surface}} = \qty{59 \pm 5}{\nano\meter}$,
consistent with ``dirty'' \ch{Nb}
(cf.\ $\lambda \approx \qty{29}{\nano\meter}$ for ``clean'' \ch{Nb}~\cite{2023-McFadden-PRA-19-044018,2023-McFadden-JAP-134-163902}).
Importantly,
for both samples we find that $\lambda_{\mathrm{surface}} > \lambda$,
implying that the screening capacity near the surface is weaker than the
average over the first \qty{\sim 150}{\nano\meter}~\cite{2023-McFadden-PRA-19-044018}.
Together,
we take this consistency as further support of inhomogeneous screening
caused by \gls{ltb} in vacuum.

In contrast,
the evidence for inhomogeneous Meissner screening in the \ch{N2} infusion sample
is less compelling.
Contrary to expectations from surface etching~\cite{2020-Checchin-APL-117-032601}
and \gls{sims}~\cite{2017-Grassellino-SST-30-094004} measurements,
our identified $L_{D}$ is large compared to the extent of near-surface nitrogen defects
and
its sizeable uncertainty hinders drawing firm conclusions about this length scale.
Similarly,
$\lambda_{\mathrm{surface}}$ was found to be close to the homogeneous $\lambda$ value,
implying that inhomogeneities in the Meissner response are either:
1) absent from this treatment;
2) beyond the resolution of the \gls{le-musr} measurements;
or
3) transpiring over depths greater than those probed in the experiments.
Note that the close agreement between screening models is
mirrored by their $\chi_{\mathrm{reduced}}^{2} < 1$,
implying that the data is overparameterized
(even in the simplest case).
Thus,
we conclude that the current data do not support the notion
of inhomogeneous screening caused by \ch{N2}
infusion;
however,
further experiments are required to be more
conclusive~\footnote{We note that there are reproducibility challenges associated with the \ch{N2} infusion ``recipe,''~\cite{2017-Grassellino-SST-30-094004} with precise control over the treatment conditions (e.g., cavity temperature, vacuum quality, furnace cleanliness, etc.) likely being crucial~\cite{2018-Dhakal-PRAB-21-032001,2022-Prudnikava-SST-35-065019}. In the future, having samples treated \emph{in situ} with \gls{srf} cavities that explicitly show \gls{hfqs} amelioration would be beneficial.}.

To further test the above findings on \gls{ltb},
we also considered an additional refinement to the Meissner profile analyses
shown in
\Cref{fig:mcfadden-2023-ltb,fig:mcfadden-2023-n2i}.
Noting that these data~\cite{2023-McFadden-PRA-19-044018} 
originate from a common batch of samples where an independent control is available
(i.e., a sample with only a ``\gls{bcp}'' treatment --- called ``baseline''
in Ref.~\citenum{2023-McFadden-PRA-19-044018}),
we re-fit the \gls{ltb} data \emph{simultaneously} using \Cref{eq:average-field,eq:london-inhomogeneous,eq:lambda-inhomogeneous},
but used the ``\gls{bcp}'' sample's screening profile to constrain $\lambda_{\mathrm{bulk}}$
(i.e., using \Cref{eq:average-field,eq:london,eq:effective-field}).
As both the sample dimensions and measurement conditions are common to these
\ch{Nb} plates~\cite{2023-McFadden-PRA-19-044018},
under the assumption that their ``true'' $d$ is also identical,
we additionally restricted the simultaneous fit so that
$B_{\mathrm{applied}}$,
$\tilde{N}$,
and
$d$
were treated as shared parameters.
The culmination of this exercise in shown in \Cref{fig:mcfadden-2023-all},
with values for the optimum parameters displayed in its inset.
Clearly,
these restrictions pose no impediment to the form of the inhomogeneous
screening in the \gls{ltb} samples,
as reflected by the good agreement of their $\lambda_{\mathrm{surface}}$
and $L_{D}$ values with those in \Cref{tab:results}.
Moreover,
the fit's common parameters (see \Cref{fig:mcfadden-2023-all}'s top panel)
are in excellent agreement with the values reported previously
for the ``\gls{bcp}'' sample~\cite{2023-McFadden-PRA-19-044018},
indicating that the fitting procedure has no adverse effect on describing its
screening behavior.
This is further confirmed by their consistency with the tabulation in
\Cref{tab:results},
apart from $\lambda_{\mathrm{bulk}}$'s \qty{\sim 2.6}{\nano\meter} difference.
This difference, albeit small, suggests that the inhomogeneous screening model
is not strongly sensitive to $\lambda_{\mathrm{bulk}}$'s absolute value,
making the literature average
(\qty{29}{\nano\meter})~\cite{2023-McFadden-PRA-19-044018,2023-McFadden-JAP-134-163902}
used when fitting the data in
\Cref{fig:mcfadden-2024,fig:mcfadden-2023-ltb,fig:mcfadden-2023-n2i}
a reasonable choice.
Overall,
these results show that having an independent means of characterizing the
``bulk'' screening properties can greatly reduce the uncertainty in quantifying
the inhomogeneous portion of the Meissner profile.

As a final remark,
we note that neither of the original \gls{le-musr} measurements~\cite{2014-Romanenko-APL-104-072601,2023-McFadden-PRA-19-044018}
were optimized for the detection of Meissner profile inhomogeneities,
instead focusing on their overall (i.e., average) form.
In future measurements,
it is apparent from \Cref{fig:mcfadden-2023-ltb,fig:mcfadden-2023-n2i,fig:mcfadden-2024}
that two regions are important to focus on:
1) the near-surface region
(i.e., $z \lesssim \qty{40}{\nano\meter}$),
where the gradual curvature of $B(z)$ is most pronounced;
and
2) deep below the surface
(i.e., $z \gtrsim \qty{120}{\nano\meter}$),
where the decay of $B(z)$ becomes close to its ``bulk'' behavior. 
While the former was suggested previously~\cite{2024-McFadden-APL-124-086101},
the latter is only evident upon inspection of the fit
curves when extrapolated beyond the current data.
Having $\langle B \rangle$ measurements in this region
($E \gtrsim \qty{25}{\kilo\electronvolt}$)
is likely crucial
for independently identifying $\lambda_{\mathrm{bulk}}$
(i.e., in the absence of a separate ``control'' sample),
as well as reducing the (rather large) uncertainty 
in the inhomogeneous screening model's fit parameters.
Similarly,
it would be beneficial to characterize the impurity content in the studied samples directly,
(e.g., using \gls{sims}~\cite{2021-Angle-JVSTB-39-024004,2022-Angle-JVSTB-40-024003}),
and
parameterize $\lambda(z)$ directly
(i.e., from their influence on the electron mean-free-path
$\ell$~\cite{1968-Goodman-JPF-29-240,1976-Schulze-ZM-67-737} --- see Ref.~\citenum{2024-Lechner-JAP-135-133902}).

\section{Summary}

In summary,
we revisited the Meissner profile data for \ch{Nb}
samples~\cite{2014-Romanenko-APL-104-072601,2023-McFadden-PRA-19-044018,2024-McFadden-APL-124-086101}
with surfaces
prepared using \gls{ltb} treatments~\cite{2004-Ciovati-JAP-96-1591,2017-Grassellino-SST-30-094004}
common to \gls{srf} cavities.
By comparing results obtained from fits to a (homogeneous) London
equation~\cite{1935-London-PRSLA-149-71}
and
the generalized London expression~\cite{1935-London-PRSLA-149-71,1981-Simon-PRB-23-4463,1986-Cave-JLTP-63-35,1994-Pambianchi-PRB-50-13659}
with an empirical model for a depth-dependent
magnetic penetration depth $\lambda(z)$~\cite{2020-Checchin-APL-117-032601},
we identify the presence of a large non-superconducting ``dead layer''
$d \gtrsim \qty{25}{\nano\meter}$ as a likely indicator for 
inhomogeneities in the near-surface Meissner response.
The inhomogeneities are most apparent for the ``\qty{120}{\celsius} bake''
treatment~\cite{2004-Ciovati-JAP-96-1591},
with the extracted length scale between ``surface'' and ``bulk'' behavior
in good agreement with both experiment~\cite{2013-Romanenko-PRSTAB-16-012001}
theory~\cite{2006-Ciovati-APL-89-022507,2024-Lechner-JAP-135-133902}. 
Conversely,
they are essentially absent for the ``\ch{N2} infusion'' treatment~\cite{2017-Grassellino-SST-30-094004},
in contrast to another report~\cite{2020-Checchin-APL-117-032601}.
Further \gls{le-musr} measurements,
covering both
shallow ($z \lesssim \qty{40}{\nano\meter}$)
and
deep ($z \gtrsim \qty{120}{\nano\meter}$)
subsurface depths,
will aid in precisely quantifying this phenomenon.

\begin{acknowledgments}
	We thank M.~Asaduzzaman and E.~M.~Lechner for useful discussions
	and a critical reading of the manuscript.
	T.~Junginger acknowledges financial support from \acrshort{nserc}.
\end{acknowledgments}

\section*{Author Declarations}

\subsection*{Conflict of Interest}
The authors have no conflicts to disclose.

\subsection*{Author Contributions}
\textbf{Ryan~M.~L.~McFadden}: Conceptualization (lead); Data Curation (lead); Formal Analysis (lead); Software (lead); Visualization (lead); Writing --- Original Draft Preparation (lead); Writing --- Review and Editing (lead).
\textbf{Tobias~Junginger}: Conceptualization (supporting); Funding Acquisition (lead); Writing --- Review and Editing (supporting).

\section*{Data Availability}
Raw data from the \gls{le-musr} measurements reported in
Refs.~\citenum{2014-Romanenko-APL-104-072601,2023-McFadden-PRA-19-044018}
were generated at the \gls{sms}, \gls{psi}, Villigen, Switzerland.
Individual data files
and
derived data supporting the findings of this work are
available from the corresponding authors upon reasonable request.

\bibliography{references.bib}

\end{document}